\documentclass{article}
\usepackage[utf8]{inputenc}
\usepackage{amsfonts}
\usepackage{amsmath}
\usepackage{colonequals}
\usepackage{graphicx}
\usepackage{epigraph}
\usepackage{hyperref}
\usepackage{array,longtable}
\usepackage{listings}

\newcommand{\xhref}[2]{{\kern0.25em}\href{#2}{#1}{\kern0.25em}}

\begin{document}

\begin{center}
\mathversion{bold}
{\bf\Large From Binary Error Correcting Codes\\[1.2ex]
  to a Relation Between Maximal D=4 and D=3 Supergravities}
\mathversion{normal}

\bigskip\bigskip\medskip

\centerline{\bf Thomas Fischbacher$^1$ and Krzysztof Pilch$^2$}
\bigskip
\bigskip
\vspace{0.5ex}

\centerline{$^1$\,Google Research,}
\centerline{Brandschenkestrasse 110, 8002 Z\"urich, Switzerland}
\vspace{2ex}

\centerline{$^2$\,Department of Physics and Astronomy,}
\centerline{University of Southern California,} \centerline{Los
Angeles, CA 90089-0484, USA}

\vspace{2ex}

{\footnotesize\upshape\ttfamily tfish @ google.com, pilch @ usc.edu} \\

\end{center}

\begin{abstract}
\noindent This short note provides (TensorFlow-based) numerical
  evidence for the embedability (in the limit of a scalar parameter
  going to infinity) of the scalar potential of dyonic
  $\mathcal{N}=8,\;D=4\;SO(8)$~supergravity into the scalar potential
  of $\mathcal{N}=16,\;D=3\;SO(8)\times SO(8)$~ supergravity. One
  finds that the dyonic $\omega$-rotation gets identified with the
  compact $U(1)$~part of the $SL(2)$~factor of the
  $SL(2)\times E_{7(+7)}$~subgroup of~$E_{8(+8)}$.
\end{abstract}

\section{Claims and Insights}

This short note is accompanied by a Google Colab
notebook\footnote{%
Available
at~\href{https://github.com/google-research/google-research/tree/master/m\_theory/colab/hamming78.ipynb}{https://github.com/google-research/google-research/tree/master/m\_theory/colab/hamming78.ipynb}, and also alongside the arXiv source code of this article.
The reader can launch this via web browser by navigating to
\href{https://colab.research.google.com/}{https://colab.research.google.com/},
selecting `GitHub' as source for a new notebook, and pasting the above url.}
(based on TensorFlow~\cite{Abadi2016}) that
numerically demonstrates the validity of each of these claims:

\begin{enumerate}

\item (From \cite{Bobev:2019dik}, Eq.\ (7.5)): One can embed
  $\mathcal{M}_{14}:=(SU(1,1)/U(1))^{\times7}$ in such a way into
  $E_{7(+7)}$ that the holomorphic superpotential is in $1{:}1$
  correspondence with the code words of the 1-bit error correcting
  (7,4,3) Hamming code~\cite{hamming1950}:
 \begin{equation}
    \label{eq:hamming7}
    \begin{aligned}
    \mathcal{W}_7 := \\ \\ \\ \\
    \end{aligned}\quad 
    \begin{gathered}
         +\zeta_1\zeta_2\zeta_3\zeta_4\zeta_5\zeta_6\zeta_7\\
         + \zeta_3\zeta_5\zeta_6\zeta_7
         + \zeta_2\zeta_4\zeta_5\zeta_7
         + \zeta_2\zeta_3\zeta_4\zeta_6
         + \zeta_1\zeta_3\zeta_4\zeta_5
         + \zeta_1\zeta_4\zeta_6\zeta_7
         + \zeta_1\zeta_2\zeta_5\zeta_6
         + \zeta_1\zeta_2\zeta_3\zeta_7
      \\
          + \zeta_1\zeta_2\zeta_4   + \zeta_1\zeta_3\zeta_6   + \zeta_1\zeta_5\zeta_7
         + \zeta_2\zeta_6\zeta_7   + \zeta_2\zeta_3\zeta_5   + \zeta_3\zeta_4\zeta_7
         + \zeta_4\zeta_5\zeta_6\\
             + 1   \,.
      \end{gathered} 
       \end{equation}

\item Expanding the $(7,4,3)$~Hamming code with a parity bit to the
  self-dual $(8,4,4)$~Hamming code, we can define a corresponding
  hypothesized holomorphic superpotential as follows, by adding 
  a factor~$\zeta_8$ to those summands that have an odd number
  of~$\zeta$-factors:
  \begin{equation}
    \label{eq:hamming8}
    \begin{aligned}
      \mathcal{W}_8 :=\\ \\ \\ \\ 
      \end{aligned}\quad 
      \begin{gathered}
      +\zeta_1\zeta_2\zeta_3\zeta_4\zeta_5\zeta_6\zeta_7\zeta_8  \\
       +\zeta_3\zeta_5\zeta_6\zeta_7
       +\zeta_2\zeta_4\zeta_5\zeta_7
       +\zeta_2\zeta_3\zeta_4\zeta_6
       +\zeta_1\zeta_3\zeta_4\zeta_5
       +\zeta_1\zeta_4\zeta_6\zeta_7
       +\zeta_1\zeta_2\zeta_5\zeta_6
       +\zeta_1\zeta_2\zeta_3\zeta_7
      \\
       +\zeta_1\zeta_2\zeta_4\zeta_8 +\zeta_1\zeta_3\zeta_6\zeta_8
       +\zeta_1\zeta_5\zeta_7\zeta_8 +\zeta_2\zeta_6\zeta_7\zeta_8
       +\zeta_2\zeta_3\zeta_5\zeta_8 +\zeta_3\zeta_4\zeta_7\zeta_8
       +\zeta_4\zeta_5\zeta_6\zeta_8\\
           +1   \,.
      \end{gathered} 
  \end{equation}

  Observing that the scalar potential corresponding to such a holomorphic superpotential
  on $\left(SU(1,1)/U(1)\right)^{\times 8}$ indeed does have many
  equilibria that align nicely (after rescaling the cosmological
  constant) with equilibiria reported in~\cite{fischbacher2009many}
  for~$\mathcal{N}=16,\;D=3\;SO(8)\times SO(8)$ supergravity,
  one may conjecture that one can indeed obtain this
  ``$(8,4,4)$~Hamming code holomorphic superpotential''
  from the~$A_1$-tensor of maximal $D=3$ supergravity.
  This indeed holds -- the details can be found in appendix~\ref{app:V_from_W}.

\item \label{claim:hamming8} Starting from the commonly used roots for the $\mathfrak{e}_{8(+8)}$ algebra,
  where the~$120-8=112$ roots of the compact~$\mathfrak{spin}(16)$ subalgebra are given by
  $(\pm 1; \pm 1;0;0;0;0;0;0) + \{\text{permutations}\}$, and the
  $128$~``$\mathfrak{spin}(16)$-spinor'' roots
  corresponding to the generators used to define the scalar manifold of
  $SO(8)\times SO(8)$ supergravity~\cite{Nicolai:2000sc,Nicolai:2001sv},
  $(\pm \frac{1}{2}; \pm \frac{1}{2};\pm \frac{1}{2};\pm \frac{1}{2};\pm \frac{1}{2};\pm \frac{1}{2};\pm \frac{1}{2};\pm \frac{1}{2})$
  (where the total number of $(-)$ signs is \emph{even}),
  it is possible to choose eight positive roots from the~$128$ such
  that when adding the corresponding eight negative roots to the set,
  no pair taken from these 16 roots have the same sign in exactly
  two positions\footnote{This would be a requirement for the
    commutator of the associated ladder operators to belong to
    $\mathfrak{spin}(16)$, but not the $\mathfrak{u}(1)^8$ generated by
    the commutators of the ladder operators for each positive root and
    its associated negative root.}.
  For any such choice, adding the corresponding negative roots,
  and encoding a $(+)$-sign as 1 and a $(-)$-sign as 0 (or vice versa) gives us
  sixteen eight-bit code words that correspond to a self-dual $(8, 4, 4)$ Hamming
  code.\footnote{A related well-known observation is that scaling the self-dual $E_8$
  lattice to integer coordinates and then taking coordinates modulo 2 yields the $(8, 4, 4)$
  self-dual Hamming code. Doing the same for the $E_7$ root lattice yields the
  $(7, 3, 4)$ `little Hamming code', while doing this for the dual $E_7$ weight
  lattice ($E_7^*$) yields the $(7, 4, 3)$ Hamming code, see e.g.~\cite{conway2013sphere,belitz2011applications}.}
  These sixteen roots then correspond to a $\mathfrak{sl}(2)^{\times8}$ subalgebra
  of~$\mathfrak{e}_8$.

\item Performing~$\omega$-deformation~\cite{dallagata2012vacua,dall2012evidence,deWit:2013ija,Dall_Agata_2014} of
  $\mathcal{N}=8,\,D=4\,SO(8)$ supergravity~\cite{deWit:1982bul},
  the superpotential in Eq.~(\ref{eq:hamming7}) acquires phase factors
  $\phi:=\exp(-i\omega)$ on summands with an \emph{odd} number
  of~$\zeta$-factors and $\bar\phi=\exp(+i\omega)$ on summands with an \emph{even} number
  of~$\zeta$-factors:
  \begin{equation}
    \label{eq:hamming7c}
    \begin{aligned}
    \mathcal{W}_{7c} :=\\ \\ \\ \\
    \end{aligned}\quad
    \begin{gathered}
 +\zeta_1\zeta_2\zeta_3\zeta_4\zeta_5\zeta_6\zeta_7\phi\\
       +\zeta_3\zeta_5\zeta_6\zeta_7\bar\phi
       +\zeta_2\zeta_4\zeta_5\zeta_7\bar\phi
       +\zeta_2\zeta_3\zeta_4\zeta_6\bar\phi
       +\zeta_1\zeta_3\zeta_4\zeta_5\bar\phi
       +\zeta_1\zeta_4\zeta_6\zeta_7\bar\phi
       +\zeta_1\zeta_2\zeta_5\zeta_6\bar\phi
       +\zeta_1\zeta_2\zeta_3\zeta_7\bar\phi
      \\
       +\zeta_1\zeta_2\zeta_4\phi +\zeta_1\zeta_3\zeta_6\phi
       +\zeta_1\zeta_5\zeta_7\phi +\zeta_2\zeta_6\zeta_7\phi
       +\zeta_2\zeta_3\zeta_5\phi +\zeta_3\zeta_4\zeta_7\phi
       +\zeta_4\zeta_5\zeta_6\phi\\
           +\bar\phi   \,.
      \end{gathered}
  \end{equation}
  Observing that the scalar potential does not change if the
  superpotential gets multiplied by a complex number of magnitude~1,
  and multiplying the above expression with~$\phi$ shows
  $\bar\phi\mathcal{W}_{7c}=\mathcal{W}_{8|\zeta_8=\phi^2}$.  Indeed,
  one finds that for~$\omega=\pi/8$, the corresponding scalar
  potential on~$(SU(1,1)/U(1))^{\times7}$ has equilibria for which the
  cosmological constants closely correspond to known solutions of the
  `dyonic SO(8)' gauging with~$\omega=\pi/8$~\cite{Berman_SO8c_2021}.
  The relation between
  the scalar potentials and superpotentials is given in
  appendix~\ref{app:V_from_W}.

\item The above properties suggest that, at least
  on~$(SL(2)/U(1))^{\times7}\sim(SU(1,1)(2)/U(1))^{\times7}$, we
  might be able to retrieve the scalar potential of $D=4\,SO(8)$
  supergravity from that of~$D=3\,SO(8)\times SO(8)$ supergravity by
  taking some suitable~$\zeta_8\to 1$ limit\footnote{Given that
    the~$\zeta$ parameters are coordinates in the Poincare disc model
    of the hyperbolic plane, this is at infinite distance from the origin.}
  -- and correspondingly,
  get the scalar potential of dyonic $D=4\,SO(8)_c$ supergravity by
  taking some $\zeta_8\to \exp(i\omega)$ limit. Hence, it seems natural to
  expect that a corresponding limit may exist for the full scalar
  potential: Using the $E_{7(+7)}\times SL(2)\subset E_{8(+8)}$ embedding
  for which we
  have~${\bf 248}\mapsto({\bf 133}, {\bf 1})+({\bf 56}, {\bf 2})+({\bf 1}, {\bf 3})$,
  the~$SL(2)$ becomes the eighth~$SL(2)$ in~$E_{8}$ that commutes with the
  seven~$SL(2)$s whose noncompact directions yield~$\mathcal{M}_{14}$.
  Considering the triality-symmetric constructions
  of~$\mathfrak{e}_7=\mathfrak{spin}(8)+{\bf 35}_v+{\bf 35}_s+{\bf 35}_c$ and
  $\mathfrak{e}_8=\mathfrak{spin(8)}^L+\mathfrak{spin(8)}^R+({\bf 8}^L_v,{\bf 8}^R_v)+({\bf 8}^L_s,{\bf 8}^R_s)+({\bf 8}^L_c,{\bf 8}^R_c)$,
  it is clear how $\mathfrak{e}_7+\mathfrak{sl}(2)$ is obtained from the
  `symmetric' pieces of the decomposition of $\mathfrak{e}_8$ with respect
  to the\footnote{Given that we can apply a triality relabeling on one of the
  $\mathfrak{spin}(8)$ algebras, there is more than one way to take a diagonal.
  The relevant diagonal here does not involve a triality rotation.}
  diagonal~$\mathfrak{spin}(8)$ subalgebra of~$\mathfrak{spin}(8)^L+\mathfrak{spin}(8)^R$.

  Using the corresponding embedding of the~$\mathfrak{e}_{7(7)}+\mathfrak{sl}(2)$
  $D=4$ scalar manifold coset generators~${\bf 35}_s+{\bf 35}_c+{\bf 1}_s+{\bf 1}_c$
  (`symmetric traceless~$8\times 8$ matrices over the spinors and co-spinors
  from $\mathfrak{e}_7$ plus multiples-of-the-identity trace-parts
  from $\mathfrak{sl}(2)$') into the space of $D=4$ scalar manifold coset
  generators~$({\bf 8}^L_s,{\bf 8}^R_s)+({\bf 8}^L_c,{\bf 8}^R_c)$ via a linear
  function~$E(v_{70}, s, c): \mathbb{R}^{70+2}\to\mathbb{R}^{128}$, one finds for
  the $D=3$ scalar potential of~$SO(8)\times SO(8)$ supergravity:
  $g_{D=3}^{-2}V_{D=3}(E(\vec 0, s, 0))<0$, and for $s>0$: $|\nabla V|>0$.
  These are non-equilibrium points with negative cosmological constant.
  If we now introduce an auxiliary (helper) function\footnote{
  The factor $-6$ is for alignment with the usual normalization of the~$D=4$ scalar potential.
  }~$H:\mathbb{R}^{70+1}\to\mathbb{R}$ as:
  \begin{equation}
    H(\vec v, s):= (-6)\,\cdot\,\frac{V_{D=3}(E(\vec v, s, 0))}{V_{D=3}(E(\vec 0, s, 0))},
  \end{equation}
  then we may conjecture that $H$ is related to the $D=4$ scalar potential of~$SO(8)$
  supergravity~$g_{D=4}^{-2}V_{D=4}: \mathbb{R}^{70}\to\mathbb{R}$ via:
  \begin{equation}
    V_{D=4}(\vec v)=\lim_{s\to\infty}H(\vec v, s).
  \end{equation}
  Numerical evidence strongly supports that this hypothesis holds
  \emph{on the full 70-dimensional scalar manifold of~$\mathcal{N}=8,\,D=4\,SO(8)$ Supergravity}!

\item The generalization to dyonic-$SO(8)$ also holds. Specifically, with
  \begin{equation}
    \label{eq:correspondence}
    H_c(\vec v, s,\omega):= (-6)\,\cdot\,\frac{V_{D=3}(E(\vec v, s\cos(2\,\omega), s\sin(2\,\omega)))}{V_{D=3}(E(\vec 0, s\cos(2\,\omega), s\sin(2\,\omega)))},
  \end{equation}
  we find:
  \begin{equation}
    V_{D=4}(\vec v, \omega)=\lim_{s\to\infty}H_c(\vec v, s, \omega).
  \end{equation}
  (As one would expect from the~$\omega$-invariance of
  the~$SO(8)$-symmetric vacuum of $SO(8)$ supergravity, one actually
  finds~$V_{D=3}(E(\vec 0, s\cos(2\,\omega), s\sin(2\,\omega)))=V_{D=3}(E(\vec 0, s, 0))$,
  so the above expression, presented `in symmetric form', can be simplified.)
  Appendix~\ref{app:claim6log} shows the numerical evidence,
  verifiable by running the accompanying Google Colab notebook.
\end{enumerate}

\section{Discussion}

The maximal (32 supercharges) gauged $D=2+1$ supergravity of Nicolai
and Samtleben~\cite{Nicolai:2000sc,Nicolai:2001sv} so far has been
mostly regarded as an exotic curiosity, as to this date there is no
known way to embed it into M theory. Correspondingly, it has perhaps
not yet received as much attention as this note suggests it should
have -- given that we observe that it indeed seems to be closely
related to the $S^7$-compactification of 11-dimensional supergravity,
i.e.  the de Wit-Nicolai model -- as well as the dyonic deformations
of that model~\cite{dall2012evidence}, for which there
currently is no known way to embed these into M theory,
either~\cite{lee2015new,Inverso_2017}. Naturally, this then means that
taking the limit in a different way will also allow us to retrieve
scalar potentials of other gaugings with already-known M theory
embeddings, such as that of `dyonic ISO(7)
supergravity'~\cite{Guarino_2015,Guarino_2016}.

\subsection{Early Clues}

As it is often useful to understand the intuition that underlies an
idea, it may be appropriate to list some major clues that contributed
to generating the idea of exploring the final claim in the list
presented above. In chronological order, these clues were:

\begin{itemize}
\item The (stable and also unstable) equilibria of maximal
  supergravities often have remarkable similarities across different
  dimensions. Notably, this also holds in particular for~$D=4$ and $D=3$.
  For example, whereas maximal $D=5$ supergravity has a
  $SU(2)\times U(1)\;\mathcal{N}=2$
  vacuum, maximal $D=4$ supergravity has a~$SU(3)\times U(1)\;\mathcal{N}=2$
  vacuum; in~$D=4$, we see a $G_2\;\mathcal{N}=1$ vacuum,
  whereas in~$D=3$, we find $G_2\times G_2\;\mathcal{N}=(1,1)$, etc. --
  see~\cite{Khavaev:1998fb,warner1983some,fischbacher2009many}.
\item As the problem of finding equilibria can be expressed entirely
  in terms of geometric invariants, the relevant properties of the
  equilibria can be expressed in terms of algebraic numbers.
  There is a general tendency for the~$D=3$ expressions to often have
  remarkably low algebraic complexity
  (see e.g.~\cite{Fischbacher:2002fx}), just as if~$D=4$ had
  to rebalance terms to make up for some loss of a more fundamental
  symmetry.
\item John Baez's article about triality and the exceptional
  groups~\cite{baez:e8} clearly was inspirational for structuring
  the code that does calculations in $E_7$ in such a way that it
  emphasizes the role of triality, despite virtually all of the other
  literature only using (anti)self-dual four-form language for $E_7$.
\item Closely studying the long list of equilibria of~$SO(8)$
  supergravity~\cite{Comsa:2019rcz} reveals some remarkable coincidences,
  such as the existence of a triplet of equilibria with residual symmetry
  $SO(4)$ where embeddings of~$SO(4)$ into~$SO(8)$ are related by
  triality. Likewise, there are closely-related-via-triality pairs
  of solutions, such as
  \xhref{S0668704}{https://arxiv.org/src/1906.00207v4/anc/extrema/S0668740/physics.pdf}--\xhref{S0698771}{https://arxiv.org/src/1906.00207v4/anc/extrema/S0698771/physics.pdf},
  \xhref{S0869596}{https://arxiv.org/src/1906.00207v4/anc/extrema/S0869596/physics.pdf}--\xhref{S0983994}{https://arxiv.org/src/1906.00207v4/anc/extrema/S0983994/physics.pdf},
  \xhref{S1068971}{https://arxiv.org/src/1906.00207v4/anc/extrema/S1068971/physics.pdf}--\xhref{S1301601}{https://arxiv.org/src/1906.00207v4/anc/extrema/S1301601/physics.pdf},
  etc., that are related by triality (see
  also~\cite{Borghese:2013dja}, as well as~\cite{fischbacher2010new}).
\item There have been various earlier indications that the 7-bit
  Hamming code is useful to understand some nontrivial aspects of M theory~\cite{Borsten_2012,Gunaydin:2020ric}.
\end{itemize}

\subsection{Outlook}

It certainly is bemusing to observe how intuition related to binary
error correcting codes did provide a relevant clue here towards
uncovering a relation between $D=4$ and $D=3$ supergravities --
especially with a view on Wheeler's ``it from bit''
essay~\cite{Wheeler:1989ftm} which proposes an agenda that includes
``\emph{[Translating] the quantum versions of string theory and of
Einstein's geometrodynamics from the language of continuum to the
language of bits}''.  One may wonder whether there are more
interesting insights that could be obtained by focusing on the
relation between remarkable lattices and binary codes -- noting
however that the (even unimodular Lorentzian) $E_{10}$ root
lattice~\cite{Gebert:e10} does not directly correspond to an error
correcting binary code -- likely due to the implicit notion of
`Euclidean distance' in the definition of error-correcting codes. This
might, however, be fixable, and suggests that a study of the relation
between~$\mathfrak{e}_{10}$ and generalized binary codes might bear
fruit.

While our focus here was exclusively on the scalar potential, this is
of course closely linked to the entire structure of the model
supersymmetry. Nominally, we are here observing a correspondence
between $D=3$ and $D=4$ supergravity in some ``AdS radius goes to
zero'' (i.e. $g^{-2}V\to-\infty$) limit. To do this, we had to ad-hoc
fix one scalar parameter and move it towards infinity without
Supergravity offering a mechanism to stabilize this configuration. We
may, at this point, only speculate whether M theory also in this
setting ``fights against being squeezed'' by growing new spatial
dimensions via some tower of massive excitations (which would mean:
degrees of freedom not present in the supergravity truncation)
collapsing to zero mass. Given our current understanding of M theory,
this speculation is however too outlandish to be taken seriously.

More tangibly, the observation that there is a~$SO(8)$ subgroup
of~$E_{8(8)}$ that rotates the eight commuting~$SL(2)$s may provide
useful to extract additional information about the structure of
the $D=4$ potential, given that this~$SO(8)$ cannot be a subgroup
of~$E_{7}$ (since it mixes the seven $U(1)$s sitting inside $E_7$
with the one outside). This might lead to an explanation for some
observations about the equilibria of the~$D=4$ scalar potential
that are currently hard to explain, such as the high degeneracies
in the mass spectra of the
equilibrium~\xhref{S1800000}{https://arxiv.org/src/1906.00207v4/anc/extrema/S1800000/physics.pdf}
\emph{despite complete breaking of~$SO(8)$} with zero residual
symmetry -- neither Lie nor discrete. Signs of a hidden~$E_{8(+8)}$
symmetry in maximal~$D=4$ supergravity are, of course,
not new~(e.g.~\cite{Ananth:2017nbi}),
and so the hope is that the rather concrete new puzzle piece
explained in this work will lead to new angles of attack to
resolve the question about the underlying symmetries of M theory.

\vskip2em
\noindent
\noindent{\bf Acknowledgments}

Thomas Fischbacher would like to thank Moritz Firsching for
independently confirming claim 3, and also Jyrki Alakuijala, George
Toderici, Ashok Popat, Rahul Sukthankar, Jay Yagnik, and Jeff Dean for
on-going support and encouragement of research that comprises a
unusual but scientifically successful applications of TensorFlow. We
also would like to thank Gianluca Inverso, David Berman, Nikolay
Bobev, Fridrik Freyr Gautason, and Hermann Nicolai for useful
discussions. Krzysztof Pilch is supported in part by DOE grant
DE-SC0011687.

\appendix

\vskip3em

{\Large\noindent {\bf Appendix}}

\vskip1em

\renewcommand{\theequation}{\Alph{section}.\arabic{equation}}
\renewcommand{\thesection}{\Alph{section}}

\section{Notebook transcript from checking claim 6} \label{app:claim6log}

It certainly is gratifying to look at the numbers that substantiate
the claim in Eq.~(\ref{eq:correspondence}).
Below, we see what happens when one randomly picks ten
(generic, non-equilibrium) points on the $E_{7(+7)}/(SU(8)/\mathbb{Z}_2)$
manifold and, in~$E_{8(8)}$, rotates outwards using~$SL(2)$,
both for~$\omega=0$ and some generic~$\omega$.

\begin{lstlisting}[language=Python,basicstyle=\scriptsize]
    >>> check_so8c_limit(omega=0, r=3, num_spot_checks=10)
    V_so8c =  -10.4047666737, V_so8xso8 =  -10.4056327510, rel_delta = 0.000083
    V_so8c =   -9.4520901976, V_so8xso8 =   -9.4528652474, rel_delta = 0.000082
    V_so8c =   -5.0839620894, V_so8xso8 =   -5.0848796316, rel_delta = 0.000180
    V_so8c =  +10.8347526006, V_so8xso8 =  +10.8334767696, rel_delta = 0.000118
    V_so8c =  -10.6441227948, V_so8xso8 =  -10.6448297935, rel_delta = 0.000066
    V_so8c =   -7.1489366176, V_so8xso8 =   -7.1496814995, rel_delta = 0.000104
    V_so8c =   +1.3503469257, V_so8xso8 =   +1.3489888355, rel_delta = 0.001006
    V_so8c =  +26.0683069298, V_so8xso8 =  +26.0658295953, rel_delta = 0.000095
    V_so8c =   +9.4490606160, V_so8xso8 =   +9.4480316002, rel_delta = 0.000109
    V_so8c =   +9.7993167162, V_so8xso8 =   +9.7976288261, rel_delta = 0.000172
    
    >>> check_so8c_limit(omega=0, r=3.5, num_spot_checks=10)
    V_so8c =  -10.4047666737, V_so8xso8 =  -10.4048838868, rel_delta = 0.000011
    V_so8c =   -9.4520901976, V_so8xso8 =   -9.4521950911, rel_delta = 0.000011
    V_so8c =   -5.0839620894, V_so8xso8 =   -5.0840862675, rel_delta = 0.000024
    V_so8c =  +10.8347526006, V_so8xso8 =  +10.8345799326, rel_delta = 0.000016
    V_so8c =  -10.6441227948, V_so8xso8 =  -10.6442184784, rel_delta = 0.000009
    V_so8c =   -7.1489366176, V_so8xso8 =   -7.1490374282, rel_delta = 0.000014
    V_so8c =   +1.3503469257, V_so8xso8 =   +1.3501631249, rel_delta = 0.000136
    V_so8c =  +26.0683069298, V_so8xso8 =  +26.0679716529, rel_delta = 0.000013
    V_so8c =   +9.4490606160, V_so8xso8 =   +9.4489213515, rel_delta = 0.000015
    V_so8c =   +9.7993167162, V_so8xso8 =   +9.7990882811, rel_delta = 0.000023
    
    >>> check_so8c_limit(omega=-4.567, r=3.0, num_spot_checks=10, scale=0.21)
    V_so8c =  +62.8771050816, V_so8xso8 =  +62.8744930759, rel_delta = 0.000042
    V_so8c =   +4.4054319480, V_so8xso8 =   +4.4041637306, rel_delta = 0.000288
    V_so8c =  +11.2968023928, V_so8xso8 =  +11.2953240524, rel_delta = 0.000131
    V_so8c =  +15.7005232999, V_so8xso8 =  +15.6989137243, rel_delta = 0.000103
    V_so8c =   +3.2290683270, V_so8xso8 =   +3.2278830222, rel_delta = 0.000367
    V_so8c =   -3.3373855762, V_so8xso8 =   -3.3383922744, rel_delta = 0.000302
    V_so8c =  +28.9523837775, V_so8xso8 =  +28.9501294956, rel_delta = 0.000078
    V_so8c =  +90.8875944078, V_so8xso8 =  +90.8832244306, rel_delta = 0.000048
    V_so8c =   -5.0578665526, V_so8xso8 =   -5.0587664349, rel_delta = 0.000178
    V_so8c =   +5.9928537431, V_so8xso8 =   +5.9908625218, rel_delta = 0.000332
    
    >>> check_so8c_limit(omega=-4.567, r=3.5, num_spot_checks=10, scale=0.21)
    V_so8c =  +62.8771050816, V_so8xso8 =  +62.8767515791, rel_delta = 0.000006
    V_so8c =   +4.4054319480, V_so8xso8 =   +4.4052603105, rel_delta = 0.000039
    V_so8c =  +11.2968023928, V_so8xso8 =  +11.2966023177, rel_delta = 0.000018
    V_so8c =  +15.7005232999, V_so8xso8 =  +15.7003054637, rel_delta = 0.000014
    V_so8c =   +3.2290683270, V_so8xso8 =   +3.2289079107, rel_delta = 0.000050
    V_so8c =   -3.3373855762, V_so8xso8 =   -3.3375218204, rel_delta = 0.000041
    V_so8c =  +28.9523837775, V_so8xso8 =  +28.9520786883, rel_delta = 0.000011
    V_so8c =  +90.8875944078, V_so8xso8 =  +90.8870029854, rel_delta = 0.000007
    V_so8c =   -5.0578665526, V_so8xso8 =   -5.0579883407, rel_delta = 0.000024
    V_so8c =   +5.9928537431, V_so8xso8 =   +5.9925842557, rel_delta = 0.000045
    
    # 'Farther out':
    >>> check_so8c_limit(omega=-4.567, r=4.0, num_spot_checks=10, scale=0.21)
    V_so8c =  +62.8771050816, V_so8xso8 =  +62.8770572401, rel_delta = 0.000001
    V_so8c =   +4.4054319480, V_so8xso8 =   +4.4054087194, rel_delta = 0.000005
    V_so8c =  +11.2968023928, V_so8xso8 =  +11.2967753155, rel_delta = 0.000002
    V_so8c =  +15.7005232999, V_so8xso8 =  +15.7004938190, rel_delta = 0.000002
    V_so8c =   +3.2290683270, V_so8xso8 =   +3.2290466170, rel_delta = 0.000007
    V_so8c =   -3.3373855762, V_so8xso8 =   -3.3374040149, rel_delta = 0.000006
    V_so8c =  +28.9523837775, V_so8xso8 =  +28.9523424881, rel_delta = 0.000001
    V_so8c =  +90.8875944078, V_so8xso8 =  +90.8875143673, rel_delta = 0.000001
    V_so8c =   -5.0578665526, V_so8xso8 =   -5.0578830348, rel_delta = 0.000003
    V_so8c =   +5.9928537431, V_so8xso8 =   +5.9928172719, rel_delta = 0.000006
\end{lstlisting}

\section{Scalar Potentials from Superpotentials} \label{app:V_from_W}

While numerics currently often appears to be the most powerful tool to
study the scalar potentials of maximal $D=4,5,6$~supergravities on the
\emph{full} coset manifolds~$E_{d(+d)}/\mathcal{K}(E_{d(+d)})$,
consistent truncation to maximal sets of commuting~$SU(1,1)\simeq
SL(2)$ subgroups yields analytically rather manageable expressions on
these low-dimensional subspaces\footnote{This likely may be a useful
starting point for explorations of larger subspaces, observing that
the Fano plane also shows in the decomposition of~$E_7$,
respectively~$E_8$, into irreducible representations of~$SL(2)^{\times
 7,8}$.}.

Following the conventions of \cite{Bobev:2019dik}, we start from the
K\"ahler potentials for the product manifold of seven, respectively
eight, Poincare discs:
\begin{equation}
  \mathcal{K}^{(7,8)} = -\sum_{j=1}^{7,\;\text{resp.}\;8}\log(1-\zeta_j\bar\zeta_j).
\end{equation}
From this, we obtain the K\"ahler metric and its inverse:
\begin{equation}
  \mathcal{K}_{a\bar b}:=\partial_a\partial_{\bar b} \mathcal{K},\quad \mathcal{K}^{a\bar b}=\left(\mathcal{K}_{a\bar b}\right)^{-1}.
\end{equation}
With the covariant derivative being given by
\begin{equation}
  \nabla_a(\cdot)=\partial_a(\cdot)+(\cdot)\partial_a\mathcal{K},
\end{equation}
the scalar potential of~$\mathcal{N}=8,\;D=4\;SO(8)$ on~$\mathcal M_{14}=(SU(1,1)/U(1))^{\times 7}$ is given by
\begin{equation}
  V_{D=4|\mathcal{M}_{14}}=2\exp(\mathcal{K})\,\left(\mathcal{K}^{a\bar b}\nabla_a\mathcal{W}_7\nabla_{\bar b}\overline{\mathcal{W}_7}-3\,\mathcal{W}_7\overline{\mathcal{W}_7}\right),
\end{equation}
while the scalar potential of~$\mathcal{N}=16,\;D=3\;SO(8)\times SO(8)$ supergravity on~$\mathcal M_{16}:=(SU(1,1)/U(1))^{\times 8}$ is found to match
\begin{equation}
  V_{D=3|\mathcal{M}_{16}}=2\exp(\mathcal{K})\,\left(\mathcal{K}^{a\bar b}\nabla_a\mathcal{W}_8\nabla_{\bar b}\overline{\mathcal{W}_8}-4\,\mathcal{W}_8\overline{\mathcal{W}_8}\right).
\end{equation}

In both cases, the superpotential can be read off from
the~$A_1$-tensor of the model: For~$D=4$, there is a 8-vector~$X^i$
such
that~$A^1_{ij}X^iX^j\cdot\prod_k\left(1-\zeta_k\bar\zeta_k\right)^{1/2}=\mathcal{W}_7$,
and in~$D=3$, there is a 16-vector~$Y^I$ such that
$A^1_{IJ}Y^IY^J\cdot\prod_k\left(1-\zeta_k\bar\zeta_k\right)^{1/2}=\mathcal{W}_8$.

\bibliographystyle{JHEP}
\bibliography{sugra}

\end{document}